\documentclass{PoS}

\title{Bounding the Higgs width at the LHC}

\ShortTitle{Bounding the Higgs width}

\author{John M. Campbell\\
        Fermilab, PO Box 500, Batavia, IL 60510\\
        E-mail: \email{johnmc@fnal.gov}}
\author{\speaker{R. Keith Ellis}\\
        Fermilab, PO Box 500, Batavia, IL 60510\\
        E-mail: \email{ellis@fnal.gov}}
\author{Ciaran Williams\\
        Niels Bohr International Academy and Discovery Center, The Niels Bohr Institute \\
Blegdamsvej 17, DK-2100 Copenhagen \O \\ 
        E-mail: \email{ciaran@nbi.dk}}

\abstract{We present results for the Standard model description of the four-lepton 
production, mediated both by Higgs boson production and by other one-loop standard model processes.
The description of four-lepton final states in MCFM v6.8 is reviewed, with special reference 
to the interference effects that can occur for identical species of leptons. We present results both for 
interference in the $\ell^- \ell^+ \ell^- \ell^+$ and in the $\ell^- \ell^+ \nu_\ell \bar{\nu}_\ell$ final state.  
Prospects for further improvement in the theoretical description of four lepton production are also reviewed.}

\FullConference{ Loops and Legs in Quantum Field Theory - LL 2014,\\
		 27 April - 2 May 2014 \\
		 Weimar, Germany, Fermilab-CONF-14-275-T}

\begin{document}

\section{Off-shell Higgs production}
\begin{figure}[b]
\begin{center} 
\includegraphics[scale=0.35,angle=270]{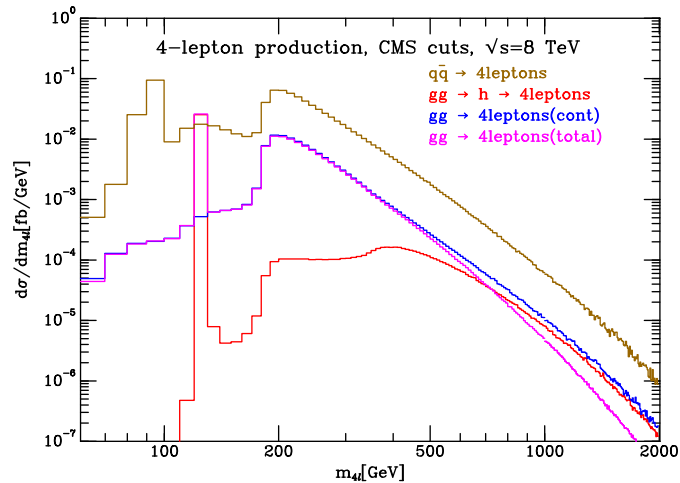} 
\caption{Overall picture at 8 TeV, (colour online). In this
figure the CMS cuts have been imposed, but the constraint $m_{4\ell}>100~{\rm GeV}$ has been 
removed to extend the range of the plot.}
\label{Bigpicture8}
\end{center}
\end{figure}

\begin{figure}
\begin{center}
\includegraphics[angle=270,width=0.5\textwidth]{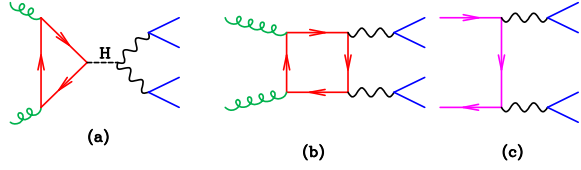}
\caption{Representative diagrams for the partonic processes considered in this paper.}
\label{alldiags}
\end{center}
\end{figure}

In an interesting recent paper, Caola and
Melnikov~\cite{Caola:2013yja} proposed to constrain the total width of
the Higgs boson using the $H \to ZZ \to 4~\mbox{lepton}$-events where
the mass of the four leptons, $m_{4l}$ is away from the Higgs
resonance region.  This method exploits the fact that at least $15\%$
of the Higgs cross section, with the Higgs boson decaying to four
charged leptons, comes from the off-peak region corresponding to a
four-lepton invariant mass above 130~GeV~\cite{Kauer:2012hd}. Various
contributions to the four charged lepton final state are shown in
Fig.~\ref{Bigpicture8}. The (red) histogram with the peak at
$m_{4l}=125$~GeV is due to the Higgs mediated diagram shown in
Fig.~\ref{alldiags}(a). Considering this diagram alone, the cross
section under the Higgs peak is proportional to $g_i^2 g_f^2/\Gamma_H$,
where $g_i$ is the coupling of the Higgs to the initial state, $g_f$
is the coupling to the final state, and $\Gamma_H$ is the total Higgs boson
width. In contrast the off-shell cross section due to this diagram is
independent of the width and proportional to $g_i^2 g_f^2$.  Thus by
comparing the measured on-peak/off-peak ratio with the theoretically
predicted ratio, one can obtain bounds on the total width of the Higgs
boson.

This simple analysis is complicated by a number of factors.  First,
the contribution to the $gg \to 4$~lepton final state, shown in
Fig.~\ref{alldiags}(b) (`the background process') interferes with the
Higgs production process. This interference is destructive at large
$m_{4 \ell}$ as it must be, since the role of the Higgs boson is to cancel
the bad high-energy behaviour of the background process. The total
rate, resulting from the total amplitude squared,
$|A^{(a)}+A^{(b)}|^2$ is shown by the (magenta) curve 
(the lowest lying curve at large $m_{4 \ell}$)
in Fig.~\ref{Bigpicture8}.  Second, the amplitude $|A^{(a)}|$ contains an
enhancement when the two constituents running in the loop are on their mass
shells.  Observation of this enhancement, which in the standard model occurs
at $m_{4 \ell} \sim 2 m_t$, would give information about the mass of the
particles that couple to the Higgs boson.  Third, the $gg$-initiated
process has to be isolated from a much bigger $q \bar{q} $ standard model
background, represented in Fig~\ref{Bigpicture8} by the uppermost (brown) curve.
The peak in this histogram at the mass of the $Z$-boson is due to singly resonant
diagrams. The curves in Fig~\ref{Bigpicture8} are taken from
ref.~\cite{Campbell:2013una} where compact analytic formula for the
background process are presented.  These formula are implemented into
the parton-level integrator MCFM, leading to numerically stable
leading-order results for the on-peak and off-peak cross sections.
The numerical results of ref.~\cite{Campbell:2013una}
correct minor numerical errors in the results of
ref.~\cite{Caola:2013yja}.

\section{MCFM:recent progress}
Before commenting on the numerical results for the four lepton
final state, we give a brief description of the advances in the
description of this process in MCFMv6.8. 
MCFM is a parton-level event integrator which gives results for a
series of processes, especially those containing the bosons $W,Z$ and
$H$ and heavy quarks, $c,b$ and $t$. The new features present in MCFMv6.8 
are:-
\begin{itemize}
\item
Identical particle effects for $ZZ \to l^- l^+ l^- l^+$ decays;
\item
Interference effects for $ZZ\to (e^- e^+) (\nu_e \bar{\nu}_e)$ and $W^+W^- \to (e^- \bar{\nu_e}) 
({\nu_e} e^+)$ decays;
\item
Inclusion of triphoton processes ($\gamma \gamma \gamma$)~\cite{Campbell:2014yka};
\item
Inclusion of Diphoton + jet processes ($\gamma \gamma$+jet).
\end{itemize}
The latter two processes are beyond the scope of this report, but we shall report on our 
results for the first two.

\subsection{Identical particle effects in $ZZ\to 4\ell$}
The original implementation of the four lepton process in MCFM was for distinct leptons, i.e. $e^- e^+ \mu^- \mu^+$,
but this has now been extended to include the case of identical charged leptons.
The interference has been included using the same method as in the POWHEG-BOX-V2~\cite{Nason:2013ydw}
to efficiently sample the more complicated phase space.

We use a set of cuts that is very similar to the ones employed in the CMS analysis~\cite{Khachatryan:2014iha}.
In order to assess the effect of the identical particle interference 
we have set the muon and electron rapidity acceptance to be the same. The detailed cuts are,
\begin{eqnarray}
&p_{T,\mu} > 7~{\rm GeV}\,, &  |\eta_{\mu}| < 2.4 \,, \nonumber \\
&p_{T,e} > 7~{\rm GeV} \,,  &  |\eta_{e}| <2.4 \,, \\
&m_{\ell\ell} > 4~{\rm GeV}\,, &m_{4\ell}>100~{\rm GeV} \,.  \nonumber
\end{eqnarray}
In addition, the transverse momentum of the hardest (next-to-hardest) lepton should
be larger than 20 (10) GeV, 
the invariant mass of the pair
of same-flavour leptons closest to the $Z$-mass should be in
the interval $40< m_{ll}<120$~GeV and the invariant mass
of the other pair should be in the interval 
$12< m_{ll}<120$~GeV. We have taken $m_H=126$~GeV and
the QCD renormalization and factorization scales
have been set equal to $m_{4\ell}/2$. The PDF set used
is CT10.

\begin{figure}
\begin{center} 
\includegraphics[scale=0.5,angle=90]{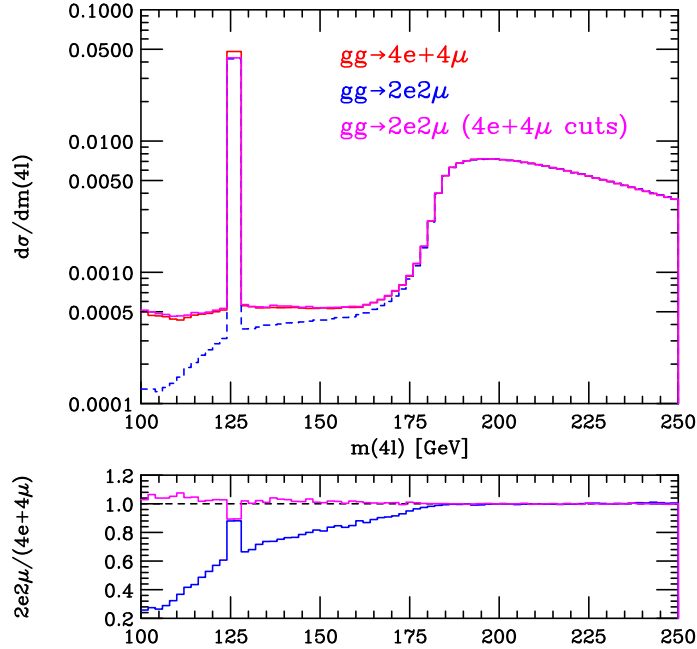} 
\caption{Interference effects in $gg \to 4\ell$, including both Higgs and
box contributions.}
\label{4l}
\end{center}
\end{figure}

We assess the interference by comparing the $m_{4\ell}$ spectrum for the $4e+4\mu$ final states
with the prediction for $2e2\mu$.  The overall normalization is then expected
to be the same, up to small differences due to the interference.  The results
are shown at $\sqrt{s}=8$~TeV for the
complete prediction, combining both Higgs and box diagrams (Fig.~\ref{4l}).
For Fig.~\ref{4l} we have computed the $2e2\mu$ process under two sets of cuts. The first
is the physical case, in which we do not attempt to form a $Z$-candidate from 
lepton pairs of different flavours, i.e. $e^- \mu^+$ and $\mu^- e^+$.  This curve is shown
in blue.  In the second case we do allow such combinations in order to make a more direct
comparison with the $4e+4\mu$ case where all leptons have the same flavor.  The resulting
curve is shown in magenta.  Comparing the curves, we see a significant difference between
the same- and different-flavor cases solely due to the fact that $Z$-candidates may be
improperly assigned in the same-flavor case.

In the bin that contains the Higgs resonance there is a significant difference
between the $4e+4\mu$ and $2e2\mu$ predictions.  There the effect of the
interference is well-known~\cite{Heinemeyer:2013tqa} and, for these cuts, it results in a
$13\%$ higher cross section for identical leptons.
There is essentially no effect from the interference in the region $m_{4\ell} > 2m_Z$.

\subsection{Interference effects for the $2\ell2\nu$ final state}
For this final state, interference can occur between final states produced by
leptonic decays of a $WW(\to \nu_l l^+ l^- \bar{\nu}_l)$ and $ZZ(\to l^- l^+ \nu_l \bar{\nu}_l)$.
It has been included using the same method as in gg2VV~\cite{Kauer:2013qba}
to efficiently sample the more complicated phase space.

We use a set of cuts that is very similar to the ones employed in the CMS analysis~\cite{CMS:xwa,CMS-PAS-HIG-13-014}.
In order to better explore the effects of the interference we have relaxed the missing $E_T$ cut
to allow for some Higgs contribution (the original analysis is aimed at a heavy Higgs bosons).
\begin{eqnarray}
&p_{T,e} > 20~{\rm GeV} \,,  &  |\eta_{e}| <2.4 \,, \\
&|m_{\ell\ell}  - m_Z| < 15~{\rm GeV}\,, &MET>25~{\rm GeV} \,.  \nonumber
\end{eqnarray}

\begin{figure}
\centering
\begin{minipage}[t]{.5\textwidth}
  \centering
  \includegraphics[width=.9\textwidth]{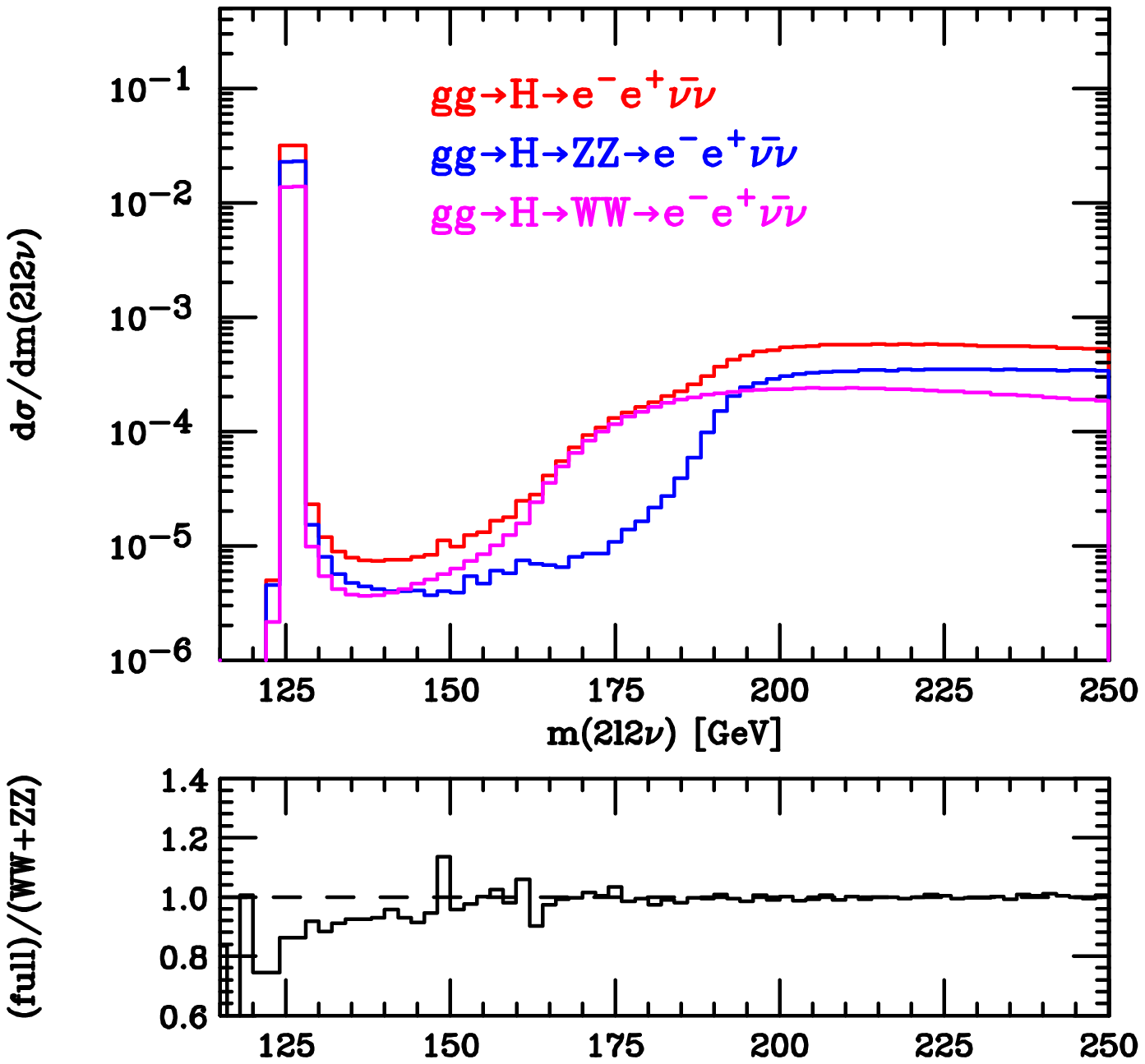}
  \caption{Interference effects in $gg \to H \to 2\ell2\nu$.}
  \label{h2l2nu}
\end{minipage}%
\begin{minipage}[t]{.5\textwidth}
  \centering
  \includegraphics[width=.9\linewidth]{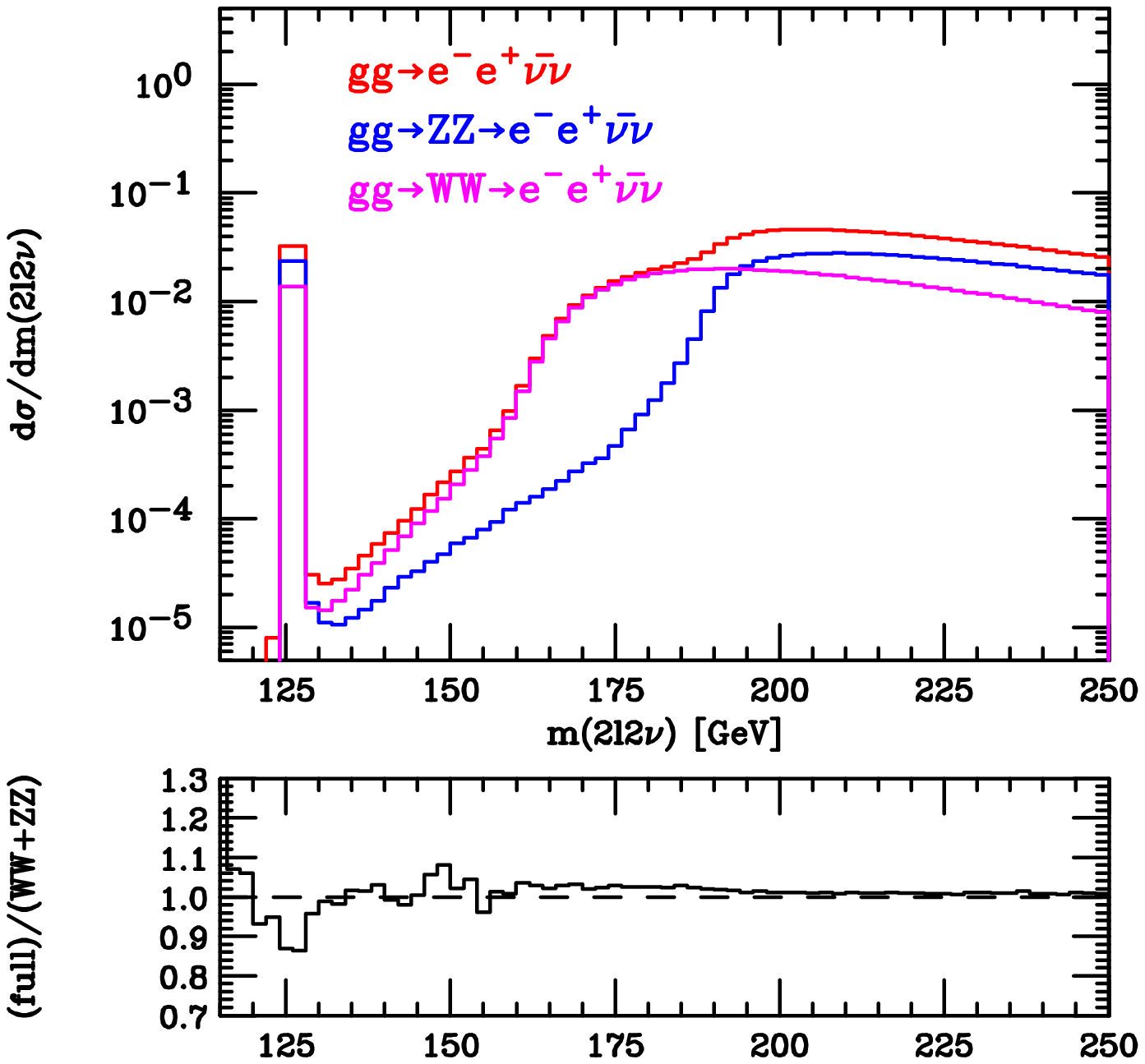}
  \caption{Interference effects in $gg \to 2\ell2\nu$, including both Higgs and
box contributions.}
  \label{2l2nu}
\end{minipage}
\end{figure}

The discussion below uses $m_{4\ell}$ for simplicity, even though it is not directly 
observable in the experiment.
We assess the interference by comparing the $m_{4\ell}$ spectrum for the sum of the $WW$ and $ZZ$ final states
with the prediction obtained by summing amplitudes before squaring.  We show results for $e^- e^+$ and a 
sum over all three neutrinos.  Note that the interference only enters for the one final state $e^- e^+ \nu_e \bar{\nu}_e$.
Results are shown at $\sqrt{s}=8$~TeV for the Higgs amplitudes alone (Fig.~\ref{h2l2nu}) and for the
complete prediction, combining both Higgs and box diagrams (Fig.~\ref{2l2nu}).
As can be seen from the lower panes in the two figures, 
there is no effect from the interference in the region $m_{4\ell} > 2m_Z$.  In contrast, there
is a signficant effect below the threshold for production of two real $Z$ bosons.  Since the
cross section is very small for the continuum diagrams alone, this is not phenomenologically
relevant except in the bin that contains the Higgs resonance.  There, the effect of the
interference is negative~\cite{Kauer:2013qba}.  For these cuts, it results in a
$25\%$ lower cross section for the Higgs diagrams alone and a $14\%$ lower cross section when
all diagrams are considered.

\section{Theoretical and experimental results in off-peak four lepton production.}
The expected cross section in the off-shell region due to the 
signal and signal-background interference is presented in ref.~\cite{Campbell:2013una}.
The number of such events expected
in the combined $7$ and $8$ TeV data sample is obtained by summing the appropriately-weighted
cross sections and normalizing to the peak cross section reported in  Ref.~\cite{CMS:xwa}. We find,
\begin{eqnarray}
N^{4\ell}_{off}(m_{4\ell}>130~\rm{GeV})
 &=& 2.78 \left( \frac{\Gamma_H}{\Gamma_H^{SM}} \right) - 5.95 \sqrt{\frac{\Gamma_H}{\Gamma_H^{SM}}} \\
N^{4\ell}_{off}(m_{4\ell}>300~\rm{GeV})
 &=& 2.02 \left( \frac{\Gamma_H}{\Gamma_H^{SM}} \right) - 2.91 \sqrt{\frac{\Gamma_H}{\Gamma_H^{SM}}}
\end{eqnarray}
The linear term is due to the Higgs contribution, whereas the square root contribution is due to 
Higgs-box diagram interference. By setting $\Gamma_H$ to its standard model value we see that the interference is
large and negative. Using the matrix elements of ref.~\cite{Campbell:2013una} the CMS collaboration
has performed a maximum likelihood fit to the on-peak and off-peak regions and 
obtain~\cite{Khachatryan:2014iha} an upper limit on the Higgs boson width of $\Gamma_H < 22$~MeV 
at a 95\% confidence level, which is 5.4 times the expected standard model value. 

\subsection{Theoretical errors}
The theoretical estimates presented so far for the above-peak production 
mediated by Higgs boson diagrams, are based on a one-loop calculation,
which in this case constitutes a leading order calculation. 
We have detailed information about the higher order corrections to the signal process, $gg \to H \to ZZ$. 
It is known that NLO corrections increase the lowest order prediction by about $80-100\%$. 
The NNLO prediction, computed in the infinite top mass limit, increases
the cross section by a further $25\%$~\cite{Ravindran:2003um,Harlander:2002wh,Dittmaier:2011ti}. 
Information about the influence of higher order corrections to the background process 
$gg \to ZZ$ is much scarcer. In this situation the CMS collaboration~\cite{CMS-PAS-HIG-14-002}
assign a $K$-factor equal to the one used for the signal to the $gg\to ZZ$ background.
They further assign a $10\%$ theoretical systematic error to this assumption.

\begin{figure}
\begin{center} 
\includegraphics[angle=270,width=0.75\textwidth]{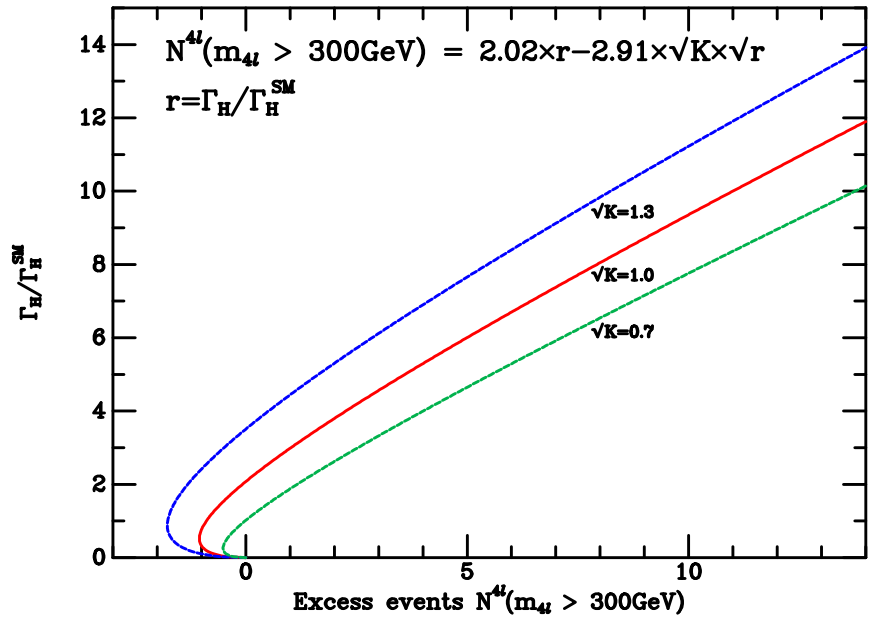} 
\caption{Influence of relative K-factor on extracted 
limit for $\Gamma_H/\Gamma_H^{SM}$ 
as a function of the number of excess events from 
$gg$-induced Higgs boson related processes.} 
\label{Kfactorbis}
\end{center}
\end{figure}

However if we allow a relative $K$-factor between the Higgs diagrams, for which the $K$-factor is known to 
be $>2$ and the continuum diagrams for which the $K$-factor is not known, the limit, assuming a bound on the 
number of excess events at $4.1$, is
\begin{equation}
\frac{\Gamma_H}{\Gamma_H^{SM}} < \{4.1,5.4,7.1\}~\mbox{for}~\sqrt{K}=\{0.7,1.0,1.3\}\, .
\end{equation}
The dependence of the assumed limit on the relative $K$-factor is shown in Fig.~\ref{Kfactorbis}.

Reference~\cite{Bonvini:2013jha} estimates the $K$-factor for the continuum process $gg \to W^+ W^-$
using the soft collinear approximation. This should be particularly appropriate for 
the case of Heavy Higgs boson, considered in ref.~\cite{Bonvini:2013jha}. Although the full soft approximation 
is not known, the missing piece can be estimated using the results for $gg \to HH$ 
which is known~\cite{Dawson:1998py} in the heavy top limit.
This argument exploits the fact that if the vector boson production is dominated by longitudinal modes, 
the equivalence theorem should allow the missing soft term to be calculated.

\begin{figure}
\begin{center} 
\includegraphics[angle=270,width=0.75\textwidth]{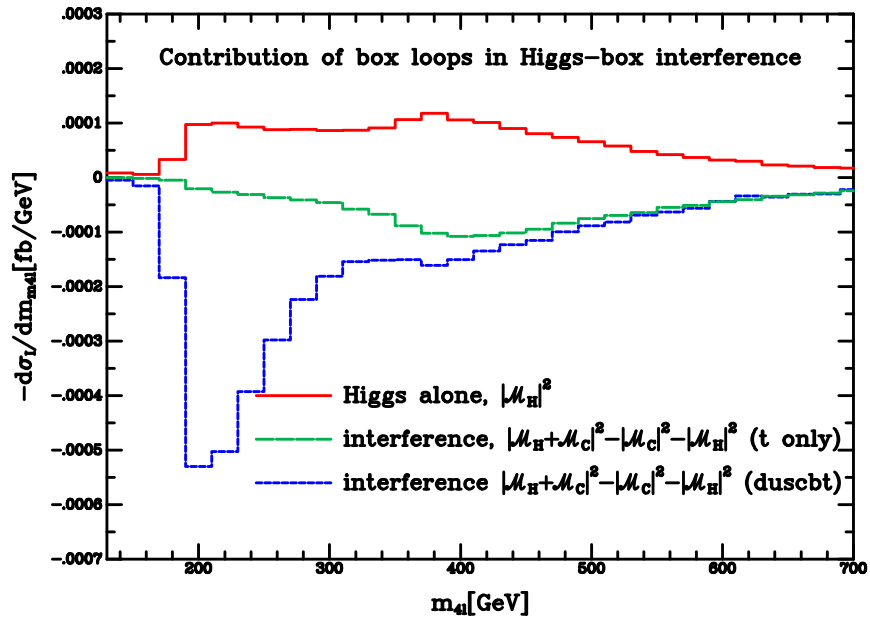} 
\caption{Contribution of the various box loops to the interference.}
\label{interferencetop}
\end{center}
\end{figure}

Figure \ref{interferencetop} shows the contribution from the various quarks to the Higgs-Box interference. 
It is only above $m_{4 \ell}>400$~GeV that the top loop (and hence the longitudinal modes)
can be said to dominate. Therefore is is only in this region that the argument of 
ref.~\cite{Bonvini:2013jha} can be strictly applied.
Moreover the discussion of ref.~\cite{Bonvini:2013jha} refers to a heavy Higgs boson for which the
interference is small and positive, whereas for the physical Higgs boson the interference is large and negative. 
It is therefore our opinion 
that the assertion that the relative $K$-factor between signal and background is small,
and therefore that the theoretical error is as small as $10\%$ (as used in the experimental paper~\cite{Khachatryan:2014iha})
is not yet established.
In the circumstances it is appropriate to perform a full NLO calculation of this interference.

\bibliographystyle{JHEP}
\bibliography{RKELandL}

\end{document}